\documentclass[a4paper,12pt]{article}

\usepackage[english]{babel}
\usepackage[utf8x]{inputenc}
\usepackage[T1]{fontenc}
\usepackage{authblk}
\usepackage[dvipdfmx]{graphicx}
\usepackage{natbib}
\usepackage{here}
\usepackage{setspace}
\usepackage{multirow}
\usepackage{lmodern}
\usepackage{amsmath, amssymb}
\usepackage{bm}
\usepackage{pdflscape}
\usepackage{authblk}
\usepackage{color}
\usepackage[table]{xcolor}

\usepackage{soul}

\usepackage{booktabs}
\usepackage{afterpage}
\usepackage{textcomp}
\usepackage{float}
\usepackage{enumerate}
\usepackage{rotating}

\definecolor{babyblue}{rgb}{0.54, 0.81, 0.94}
\definecolor{babypink}{rgb}{0.96, 0.76, 0.76}

\usepackage[margin=1in]{geometry}

\usepackage{amsmath}
\usepackage{graphicx}
\usepackage[colorlinks=true, allcolors=blue]{hyperref}

\usepackage{caption}
\usepackage{subcaption}

\usepackage{lineno}

\begin{document}

    \begin{center}
        \vspace*{1cm}
        \large
        \textbf{Trends in bird abundance differ among protected forests but not bird guilds}\\
         \normalsize
           \vspace{5mm}
         Jeffrey W. Doser\textsuperscript{1, 2}, Aaron S. Weed\textsuperscript{3}, Elise F. Zipkin\textsuperscript{2, 4}, Kathryn M. Miller\textsuperscript{5}, \\ Andrew O. Finley\textsuperscript{1, 2, 6}
         \vspace{5mm}
    \end{center}
    \small
         \textsuperscript{1}Department of Forestry, Michigan State University, East Lansing, MI, 48824, USA \\
         \textsuperscript{2}Ecology, Evolution, and Behavior Program, Michigan State University, East Lansing, MI, 48824, USA \\
         \textsuperscript{3}Northeast Temperate Inventory and Monitoring Network, National Park Service, Woodstock, VT 05091, USA \\
         \textsuperscript{4}Department of Integrative Biology, Michigan State University, East Lansing, MI, 48824, USA \\
         \textsuperscript{5}Northeast Temperate Inventory and Monitoring Network, National Park Service, Bar Harbor, ME, 04609, USA \\
         \textsuperscript{6}Department of Geography, Environment, and Spatial Sciences, Michigan State University, East Lansing, MI, 48824, USA \\
          \noindent \textbf{Corresponding Author}: Jeffrey W. Doser, telephone: (585) 683-4170; email: doserjef@msu.edu; ORCID ID: 0000-0002-8950-9895 \\
          \textbf{Running Head}: Trends in forest bird abundance
         \doublespacing
         
   \newpage
   \section*{Abstract}
 
Improved monitoring and associated inferential tools to efficiently identify declining bird populations, particularly of rare or sparsely distributed species, is key to informed conservation and management across large spatio-temporal regions. We assess abundance trends for 106 bird species in a network of eight forested national parks located within the northeast U.S.A. from 2006-2019 using a novel hierarchical model. We develop a multi-species, multi-region removal sampling model that shares information across species and parks to enable inference on rare species and sparsely sampled parks and to evaluate the effects of local forest structure. Trends in bird abundance over time varied widely across parks, but species showed similar trends within parks. Three parks (Acadia National Park and Marsh-Billings-Rockefeller and Morristown National Historical Parks (NHP)) decreased in bird abundance across all species, while three parks (Saratoga NHP and Roosevelt-Vanderbilt and Weir-Farm National Historic Sites) increased in abundance. Bird abundance peaked at medium levels of basal area and high levels of percent forest and forest regeneration, with percent forest having the largest effect. Variation in these effects across parks could be a result of differences in forest structural stage and diversity. By sharing information across both communities and parks, our novel hierarchical model enables uncertainty-quantified estimates of abundance across multiple geographical (i.e., network, park) and taxonomic (i.e., community, guild, species) levels over a large spatio-temporal region. We found large variation in abundance trends across parks but not across bird guilds, suggesting that local forest condition might have a broad and consistent effect on the entire bird community within a given park. Research should target the three parks with overall decreasing trends in bird abundance to further identify what specific factors are driving observed declines across the bird community. Understanding how bird communities respond to local forest structure and other stressors (e.g., pest outbreaks, climate change) is crucial for informed and lasting management. \\

\noindent \textbf{Keywords}: avian populations, Bayesian analysis, community dynamics, hierarchical model, imperfect detection, National Park Service, removal sampling, species richness

\newpage

\section*{Introduction}

Developing sampling designs and efficient statistical methods to monitor trends in species (e.g., \citealt{Lany2020}) and communities (e.g., \citealt{Farr2019}) is critical to inform management of landscapes, wildlife, and other natural resources \citep{Stem2005}. Species occupancy and abundance trends have been used extensively over the last decade to identify and prioritize  management tasks needed to protect wildlife (e.g., \citealp{Guisan2005, Zipkin2010, VanderYacht2016}). Species distribution models and similar methods are key tools to evaluate changes in populations and communities (e.g., \citealt{Wilkinson2019}), an increasingly important task as climate change and habitat loss cause new challenges to conservation. Monitoring of bird species distributions in particular has received much attention in the ecological literature as a result of their popularity, wide development of statistical methods to account for imperfect detection (e.g., \citealt{Farnsworth2002, Royle2004}), large-scale public science programs (e.g., Breeding Bird Survey, eBird), and bird sensitivity  to environmental stressors \citep{bibby1999, Canterbury2000, gregory2003}. 

While monitoring changes in species distributions over time is a common task, assessments of abundance trends is less common and more difficult, but critical as declines in abundance can lead to reductions in ecosystem services \citep{Inger2015a, Rosenberg2019}. Key studies have shown large declines in bird abundance in the last several decades across extensive spatio-temporal regions in North America \citep{Stanton2018, Rosenberg2019} and Europe \citep{Inger2015a}. Such research, finding large declines in both common and rare bird species, suggests the need for large-scale monitoring of bird abundance.

Forest bird species in particular play an essential role in forest ecosystems primarily via their foraging ecology that results in ecological services such as scavenging carcasses, nutrient cycling, pest management, and seed dispersal \citep{Bruns1960, Whelan2008, whelan2010, Whelan2015}. Understanding changes in forest bird abundance over time is a key task in maintaining ecological integrity of forest ecosystems. Diversity of forest birds is strongly linked to structural forest composition and complexity \citep{macArthur1958,Macarthur1964, macArthur1961, Macarthur1966, karr1971, Willson1974}, but how that complexity and composition affects bird abundance over time is unresolved.

Monitoring of bird abundances occurs primarily through the use of point-count surveys, in which observers record all birds seen at a location for a set amount of time. Sampling and accompanying modeling techniques to address imperfect detection in count data have been widely developed over the last 50 years and include removal sampling \citep{Farnsworth2002}, distance sampling \citep{buckland1993}, repeated counts \citep{Royle2004}, capture-recapture \citep{Otis1981, Royle2009}, spatial capture-recapture \citep{efford2004, borchers2008spatially, royle2008hierarchical}, and useful combinations of such methods \citep{Amundson2014, Miller2019}. Many extensions of these modeling techniques exist to meet objectives of different types of monitoring programs, such as to estimate population size of multiple species at a given point in time (e.g., \citealt{Chandler2013, Farr2019}) and assess trends in species abundance across larger time periods \citep{Dail2011, Zhao2019}. Recently, the multi-species occupancy model \citep{Dorazio2005estimating, gelfand2005modelling, dorazio2006estimating, gelfand2006explaining} was extended to a multi-region framework that enables modeling of occupancy and species richness across independent spatial units that comprise a broader network \citep{Sutherland2016, wright2020}. An analogous approach could model abundance across a large network to provide inference on factors driving spatio-temporal changes in the abundances of common and rare species.

In 2006, the Northeast Temperate Inventory and Monitoring Network (NETN) of the National Park Service (NPS) began a volunteer monitoring program to collect data on bird community abundance and composition across upland forest habitats in eight parks. We used these data in a network-wide analysis to characterize spatio-temporal dynamics of forest bird abundance from 2006-2019. We estimated park-specific trends in bird abundance to evaluate whether patterns in the bird community change among protected areas, which may signify where and whether management efforts would be most beneficial. Within each park, we also estimated trends of functionally similar groups of birds (i.e., guilds) and of individual species to determine if management should target the whole community, specific guilds, or individual species. We also sought to understand whether bird abundance is more driven by presence of breeding habitat (i.e., percent of forest within a 1km radius) or quality of the breeding habitat characterized by the amount of live tree basal area and forest regeneration, which have been shown to influence bird species diversity \citep{flaspohler2002, Rankin2015, Zipkin2010, VanderYacht2016}. To do this, we extend the basic removal sampling model \citep{Farnsworth2002} and the multi-species removal sampling model \citep{Chandler2013} to a multi-region, multi-species framework that directly accounts for imperfect detection and shares information across species within each park and across the network of parks. This novel hierarchical approach enables estimation for rare species and estimation of parameters at parks where there is a paucity of data. 

We predicted high variability in abundance trends across parks and species, which we hypothesized would reveal which species and locations should be targeted for management interventions. We also predicted the amount of local forest cover (percent forest within a 1km radius) would have the largest (and positive) effect on forest bird abundance \citep{Ladin2016}. We expected little variation in the direction and magnitude of trends of species within a guild and larger variation among the different guilds \citep{OConnell2000}. For instance, we predicted forest regeneration would have a positive effect on ground and shrub nesting species, as the amount of regeneration is negatively correlated with the amount of deer browsing \citep{Russell2001, Augustine2003}, which has been shown to reduce the occupancy probability of ground-nesting and understory species \citep{Zipkin2010}. We expected forest interior obligate and canopy nesting species to show positive relationships with the amount of basal area \citep{Rankin2015}, while shrub nesters and forest ground nesters will show either a negative relationship with basal area or a curvilinear relationship, peaking at intermediate values \citep{flaspohler2002, Rankin2015, VanderYacht2016}.

\section*{Materials and Methods}

\subsection*{\textit{Study Site and Sampling Methods}}

\begin{figure}[!h]
    \centering
    \includegraphics[width=10cm]{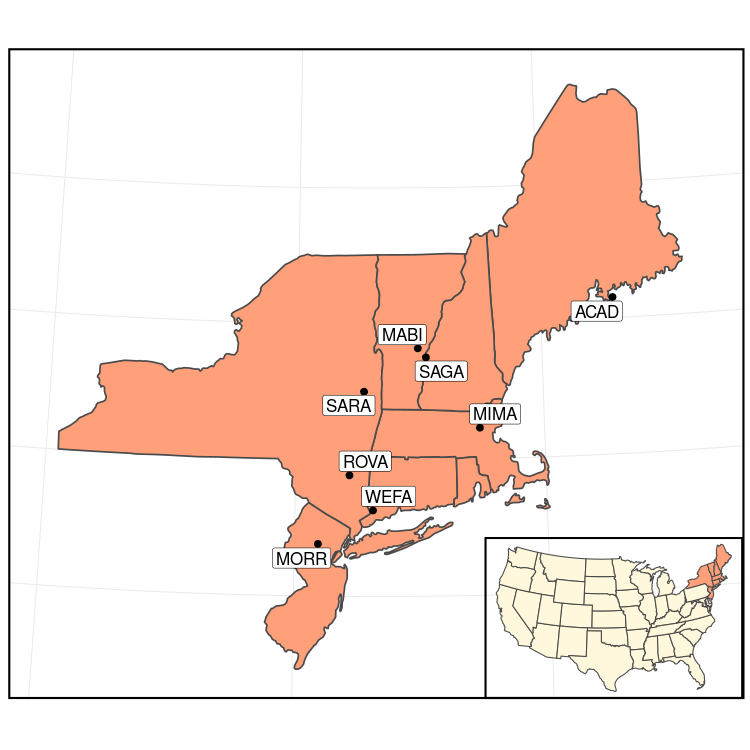}
    \caption{}
    \label{fig:studyArea}
\end{figure}

The NETN monitored eight parks across the northeastern United States of America (USA) from 2006-2019 (Figure~\ref{fig:studyArea}, \citealt{netnBird2015}). Table~\ref{tab:netnInfo} includes the park acronyms used throughout the remainder of the manuscript, the total amount of forested land in each park, and a summary of data available for parks (as sampling intensity varied across years and parks). The number of point count locations used to monitor forest birds (henceforth called points) ranged across parks from 5 (WEFA) to 51 (ACAD). The NETN established point count locations based on four criteria: 1) points were 200-250m apart to avoid duplicate sampling; 2) points were located at least 50m from forest edges; 3) points were located in the dominant forest type of each park; and 4) at least one forest vegetation sampling plot was located within 50m of at least one point in each dominant forest type. To meet these criteria, the NETN used a systematic grid sampling frame to select points at all parks except ACAD, where the NETN used the Generalized Random Tesselation Stratified algorithm \citep{Stevens1997, Stevens2004}. See \cite{netnBird2015} for further details on point selection and protocol methods.

Volunteers performed 10-minute surveys at points located along permanent forested, transects, annually during the breeding season (May-July). At each point, an observer recorded the time of day and species of individual birds they saw or heard within ten, one-minute increments. Observers recorded each individual bird only the first time they detected the bird during the point count. Observers performed point count surveys in the early morning and during adequate weather conditions, when vocalization is most likely. There was large variability in the number of points surveyed in any specific year within parks as a result of volunteer availability (Table~\ref{tab:netnInfo}). 

We used the time-interval of detections to estimate species' abundances with a model that incorporated detection probabilities and the removal sampling data \citep{Farnsworth2002, nichols2009}. We developed a model that shares information across parks and species within a hierarchical framework because data were sparse at some parks (i.e., WEFA, SAGA, SARA) and for rarely detected species. This approach enables detection corrected estimates of abundance at lesser sampled parks and for rare and/or elusive species, and also yields more precise estimates for all species at all parks, analogous to multi-species occupancy models (e.g., \citealt{Zipkin2010}).

\subsection*{\textit{Forest Covariates}}

In addition to bird monitoring stations, the NETN has over 300 forest vegetation monitoring plots implemented across the same eight parks \citep{netnForest2016}. Some of these plots are co-located within 50m of the bird point count survey points, which enables assessment of the relationship between forest structure and bird abundance. For points that did not have an associated forest plot within 50m, we used covariate data from the closest forest vegetation monitoring plot occurring in the same general forest cover type.  

We evaluated the influence of three forest covariates: amount of forest regeneration, amount of live basal area, and percent forest within a 1km radius around the survey points. We used the amount of forest regeneration (stems ha$^{-1}$) as a metric to quantify advanced tree regeneration in the forest understory, and computed regeneration as the total observed density of seedlings (diameter at breast height (DBH) < 1 cm and at least 15 cm tall) plus the total observed density of saplings (1 cm $\leq$ DBH $\leq$ 10 cm) per plot. We calculated the amount of living basal area ($\text{m}^2/\text{ha}$) for a given canopy tree in a plot from the diameter at breast height (DBH) and then summed over all species in the plot. Field crews measured these forest variables once every four years in each plot, resulting in 3-4 measurements of basal area and regeneration for a given point over the study period. We averaged these values so that each point was associated with a single value of basal area and forest regeneration that represented the average of these forest characteristics across the 14-year study period to provide a broad overview of how forest condition influences forest bird abundance. We obtained the percent forest within a 1km radius around the survey location in 2016 using the National Landcover Database \citep{Homer2015}. We standardized all covariates to have mean zero and a standard deviation of one to facilitate comparison of effect sizes and relative importance. Preliminary analyses suggested linear effects of all three variables and a quadratic effect of basal area on abundance would yield the best-fitting model. 

\subsection*{\textit{Modeling Framework}}

\begin{figure}
    \centering
    \includegraphics[width=15cm]{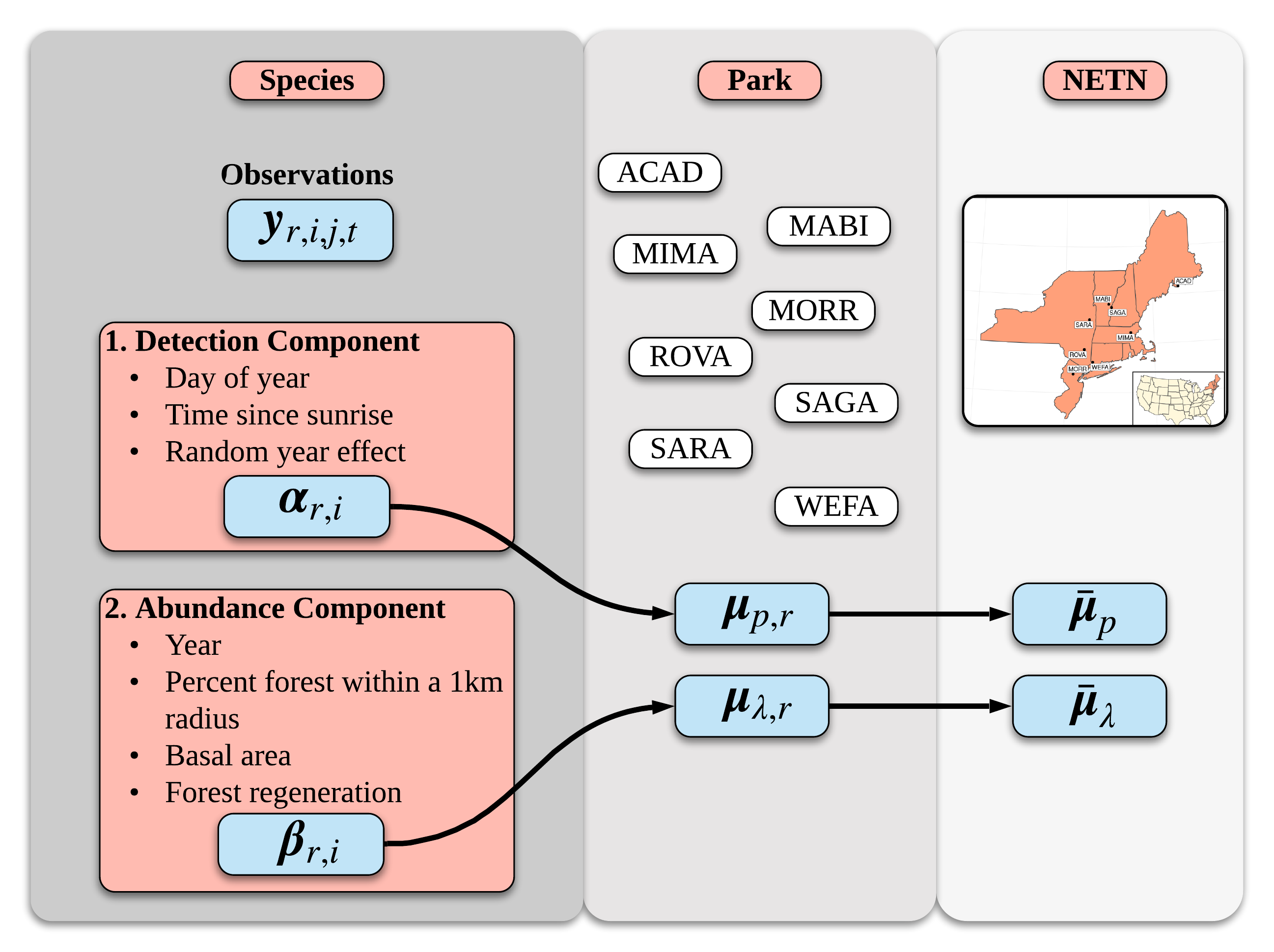}
    \caption{}
    \label{fig:modelDiagram}
\end{figure}

We developed a hierarchical model (e.g., \citealt{berliner1996, gelman04, royle2006JABES, royle2008hierarchicalbook, hooten}) to estimate trends in community abundance and individual species abundance across multiple geographically distinct parks (\citealt{Sutherland2016, wright2020}; see Figure~\ref{fig:modelDiagram} for overview of modeling framework). To share information across species and effectively model communities of species within each park, we view species-level parameters as arising from common park-level hyperparameters, which enables improved precision on species level estimates \citep{yamaura2011JAE, yamaura2012BC}. We extend this concept to another hierarchical layer to share information across the entire meta-community by drawing park-level hyperparameters from a common network-level distribution \citep{Sutherland2016, wright2020}. The model is thus a multi-level extension of a generalized linear model (GLM) with the following hierarchical levels: 1) detection model to accommodate imperfect detection for each species; 2) process model of the true latent abundance of each species; 3) park level that models species-specific parameters from a common park (i.e., community) level distribution to induce dependence among species in a park and share information across the community; and 4) network-level that views park-level parameters as arising from a common network-level distribution to share information across all parks in the network. While alternative approaches exist to model species jointly via direct dependency of species detections and/or abundances (e.g., \citealt{ovaskainen2010modeling, ovaskainen2011making, warton2015so}), the hierarchical approach we employ is a mechanistic alternative that enables inference at multiple ecological levels while accounting for imperfect detection. This approach has been shown to provide good model fit, ecological insights, and key information to support management decisions (e.g., \citealt{Zipkin2009, Chandler2013, Linden2013, zipkin2020Science}). 

Following the removal sampling protocol, our model uses the time period of first observation to account for imperfect detection of individuals during the survey by estimating a detection probability that is the product of availability and detectability \citep{Farnsworth2002, nichols2009, kery2015applied}.  We used intervals of two-minutes in length to summarize detections from the available data resulting in a total of $B = 5$ intervals in which an individual bird could be detected during the survey.

We implemented point count surveys using the removal sampling protocol at $r = 1, \dots, 8$ geographically distinct parks for $i = 1, \dots, n_r$ species at $j = 1 \dots, J_r$ points for each of $t = 1, \dots, T_r$ years.  We define the vector $\bm{y}_{r, i, j, t}$ as the number of individuals of species $i$ encountered during year $t$ at point $j$ within park $r$ in each of the $B$ time intervals. Subsequently, we define $\bm{y}^*_{r, i, j, t}$ as equivalent to $\bm{y}_{r, i, j, t}$ with an additional value representing the number of individuals that were not detected in any time interval. Following the basic removal sampling model, we view the data $\bm{y}^*_{r, i, j, t}$ as arising from a multinomial distribution with cell probabilities $\bm{\pi}^*_{r, i, j, t}$ conditioned on the latent abundance $N_{r, i, j, t}$. We then view the latent abundance $N_{r, i, j, t}$ as a Poisson distributed random variable with mean $\lambda_{r, i, j, t}$. The multinomial observation model thus allocates the total latent number of birds of species $i$ at point $j$ in park $r$ during year $t$ ($N_{r, i, j , t}$) across the $B$ removal time periods and the additional category for individuals not observed. Accordingly our models for both the sampling and biological processes, respectively, take the following forms

\begin{equation*}
\begin{split}
    \bm{y}^*_{r, i, j, t} &\sim \text{Multinomial}(N_{r, i, j, t}, \bm{\pi}^*_{r, i, j, t}) \\
    N_{r, i, j, t} &\sim \text{Poisson}(\lambda_{r, i, j, t}).
\end{split}
\end{equation*}

Given this formulation, we can use the relationship between the multinomial and Poisson distributions to  directly model $\bm{y}_{r, i, j, t}$ as a series of conditionally independent Poisson distributions by analytically deriving the marginal likelihood \citep{Royle2004ABC, Dorazio2005Biometrics, kery2015applied, Yackulic2020}. For each $b = 1, \dots , B$ removal period, our marginal likelihood is defined as  

\begin{equation*}
    y_{r, i, j, t, b} \sim \text{Poisson}(\lambda_{r, i, j, t} \cdot \pi_{r, i, j, t, b}),
\end{equation*}

where $\bm{\pi}_{r, i, j, t}$ is equivalent to $\bm{\pi}^{*}_{r, i, j, t}$ but does not include the probability of failing to detect an individual that was truly present. Using this conditional likelihood, we do not directly obtain estimates of the latent abundance $N_{r, i, j, t}$, but we can subsequently draw these values from a Poisson distribution with mean $\lambda_{r, i, j, t}$. 

\subsubsection*{\underline{Detection Model}}

Under the removal sampling protocol, the cell probability of being observed in the $b$th time interval is defined as 

\begin{equation*}
    \pi_{r, i, j, t, b} = p_{r, i, j, t}(1 - p_{r, i, j, t})^{b - 1},
\end{equation*}

where $p_{r, i ,j ,t}$ is the probability of an individual of species $i$ being detected in at least one time interval during year $t$ at point $j$ within park $r$. We incorporated covariate and random effects in the detection probability models as 

\begin{align*}
    \text{logit}(p_{r, i, j, t}) = \alpha_{0, r, i} + \alpha_{1, r, i} \cdot \text{DAY}_{r, j, t} + \alpha_{2, r, i} \cdot \text{DAY}^2_{r, j, t} + \alpha_{3, r, i} \cdot \text{TIME}_{r, j, t} + \alpha_{4, r, t},
\end{align*}

where $\alpha_{0, r, i}$ is the intercept for species $i$ in park $r$, and $\alpha_{1, r, i}, \alpha_{2, r, i},  \alpha_{3, r, i}$ are regression coefficients representing the linear effect of Julian date (DAY$_{r,j,t}$), the quadratic effect of Julian date ($\text{DAY}^2_{r, j, t}$), and the time since sunrise ($\text{TIME}_{r, j, t}$), respectively, on the detection probability of species $i$ at park $r$. We included a random year effect stratified by park, $\alpha_{4, r, t}$, to account for variability in detection probability across years and parks that follows a normal distribution with mean 0 and variance $\sigma^2_{4, p}$. The species/park specific effects $\alpha_{0, r, i}, \alpha_{1, r, i}, \alpha_{2, r, i}$, and $\alpha_{3, r, i}$ follow distributions with common park-specific parameters to enable sharing of information across species within a given park. Specifically, we model $\alpha_{1, r, i}$ as

\begin{equation*}
\alpha_{1, r, i} \sim \text{Normal}(\mu_{p, 1, r}, \sigma^2_{p, \text{REG}, 1}),
\end{equation*}

where $\mu_{p, 1, r}$ represents the mean linear effect of day on detection probability across the entire species community in park $r$, and $\sigma^2_{p, \text{REG}, 1}$ represents the variability in this effect across the different species comprising the community in park $r$. Models of $\alpha_{0, r, i}, \alpha_{2, r, i}$, and $\alpha_{3, r, i}$ are defined analogously.  The $\mu_{p, 1, r}$ coefficients provide a straightforward way to compare how covariates influence the bird communities across different parks. Further, $\mu_{p, 1, r}$ follow a common distribution to enable sharing of information across parks, which allows for estimates in sparsely sampled parks (i.e., WEFA, SAGA, SARA). To do this, each $\mu_{p, 1, r}$ is modeled as 

\begin{equation*}
    \mu_{p, 1, r} \sim \text{Normal}(\bar{\mu}_{p, 1}, \sigma^2_{p, \text{META}, 1}),
\end{equation*}

where $\bar{\mu}_{p, 1}$ is the mean linear effect of day on detection probability across all species and parks (i.e., the meta-community), and $\sigma^2_{p, \text{META}, 1}$ represents the variability in the linear effect of day on detection probability across parks within the meta-community. Models of $\mu_{p, 0, r}, \mu_{p, 2, r}$, and $\mu_{p, 3, r}$ are defined analogously.

\subsubsection*{\underline{Abundance Model}}

As with the detection model, we incorporated random effects and covariates at multiple levels of the hierarchy to explain variation in $\lambda_{r, i, j, t}$. We model $\lambda_{r, i, j, t}$ using the following specification

\begin{equation*}
\begin{split}
    \text{log}(\lambda_{r, i, j, t}) =\; &\beta_{0, r, i} + \beta_{1, r, i} \cdot \text{YEAR}_{r, t} + \beta_{2, r, i} \cdot \text{REGEN}_{r, j} + \beta_{3, r, i} \cdot \text{FOR}_{r, j} + \\
    &\beta_{4, r, i} \cdot \text{BA}_{r, j} + \beta_{5, r, i} \cdot \text{BA}^2_{r, j},
\end{split}
\end{equation*}

where $\beta_{0, r, i}$ is the abundance intercept for species $i$ in park $r$, and $\beta_{1, r, i}$, $\beta_{2, r, i}$, $\beta_{3, r, i}$, $\beta_{4, r, i}$, and $\beta_{5, r, i}$ are regression coefficients representing the linear effect of year ($\text{YEAR}_{r, t}$), regeneration ($\text{REGEN}_{r, j}$), local forest cover ($\text{FOR}_{r, j}$), basal area ($\text{BA}_{r, j}$), and quadratic effect of basal area ($\text{BA}^2_{r, j}$), respectively, on the abundance of species $i$ at park $r$. We define all species and park-specific parameters as random effects that follow a common distribution within each park. Park-specific parameters subsequently follow a common meta-community distribution. For $\beta_{1, r, i}$ we have  

\begin{align*}
    & \beta_{1, r, i} \sim \text{Normal}(\mu_{\lambda, 1, r}, \sigma^2_{\lambda, \text{REG}, 1}) \\
    & \mu_{\lambda, 1, r} \sim \text{Normal}(\bar{\mu}_{\lambda, 1}, \sigma^2_{\lambda, \text{META}, 1}),
\end{align*}

where $\mu_{\lambda, 1, r}$ is the mean year effect across the entire community in park $r$, $\sigma^2_{\lambda, \text{REG}, 1}$ is the variability of the year effect across all species in the meta-community, $\bar{\mu}_{\lambda, 1}$ is the mean year effect across the entire meta-community, and $\sigma^2_{\lambda, \text{META}, 1}$ is the variability of the park year effects across the entire meta-community. Models of $\beta_{0, r, i}, \beta_{2, r, i}, \beta_{3, r, i}, \beta_{4, r, i}$, and $\beta_{5, r, i}$ are defined analogously. 

\subsection*{\textit{Bird Guilds}}

To assess the effects of local forest structure on species' abundances, we assigned birds to behavioral and physiological response guilds following \cite{OConnell2000}. Bird guilds are groups of species that respond in similar ways to environmental changes as a result of similar uses of the environment \citep{root1967niche, Verner1984, Szaro1986}. Overall, we categorized 16 guilds in three biotic integrity elements (Functional, Compositional, Structural) as specialist or generalist guilds depending on each guild's relationship to the landscape structure and function. Specialist guilds consist of species with highly specific habitat requirements, and generalist guilds consist of species that can utilize a wide range of habitats. We selected guilds to indicate different aspects of species' life history traits, which can lead to a single species being assigned to multiple guilds (\citealt{netnBird2015}; Table~\ref{tab:guildInfo}). Based on the recommendations in \cite{Pacifici2014}, we only used bird guilds in post-hoc assessment.

\subsection*{\textit{Model Estimation}}

We implemented the model using a Bayesian framework (e.g., \citealt{berliner1996, clarkTextbook, hooten}). The Bayesian approach is particularly useful in our setting because it easily accommodates missing data and facilitates inference about derived quantities such as covariate effects across different bird guilds with propagated uncertainty across all hierarchical levels (e.g., species and parks). We specified vague normal priors for regression coefficients and vague gamma priors for variance parameters. We fit our models using Markov chain Monte Carlo (MCMC) in JAGS \citep{Plummer03jags} within the \verb+R+ statistical environment \citep{r} using the \verb+jagsUI+ \citep{jagsUI} package. We ran models for 50,000 iterations with a burn-in period of 45,000 iterations and a thinning rate of 2. We assessed model convergence using visual assessment of trace plots and the Gelman-Rubin R-hat diagnostic, where convergence was considered to occur for all values of R < 1.1 \citep{brooks1998, gelman04}. We assessed model fit using a Bayesian p-value approach \citep{Gelman1996, hooten}, which indicated a successful model fit (mean = 0.21). We interpret a parameter as ``significant'' if the 95\% credible interval does not include zero. We performed all subsequent analysis in \verb+R+ using the \verb+coda+ package \citep{coda}. All code and data are available on Zenodo \citep{dataAvail}.

\section*{Results}

We observed 106 bird species from 2006-2019 across the eight parks, with an average of 59 species observed at an individual park (Table~\ref{tab:summaryStats}). SARA had the highest observed species richness with a total of 71 observed species, while WEFA had the lowest richness, with a total of 45 observed species. Most species in each park's community were rare, as the average point-level observed abundance of a species across all parks was 0.17, with average point-level abundance highest at MORR (0.22) and lowest in ACAD (0.11). 

\subsection*{\textit{Temporal trends in bird abundance}}

\begin{figure}
    \centering
    \includegraphics[width = 15cm]{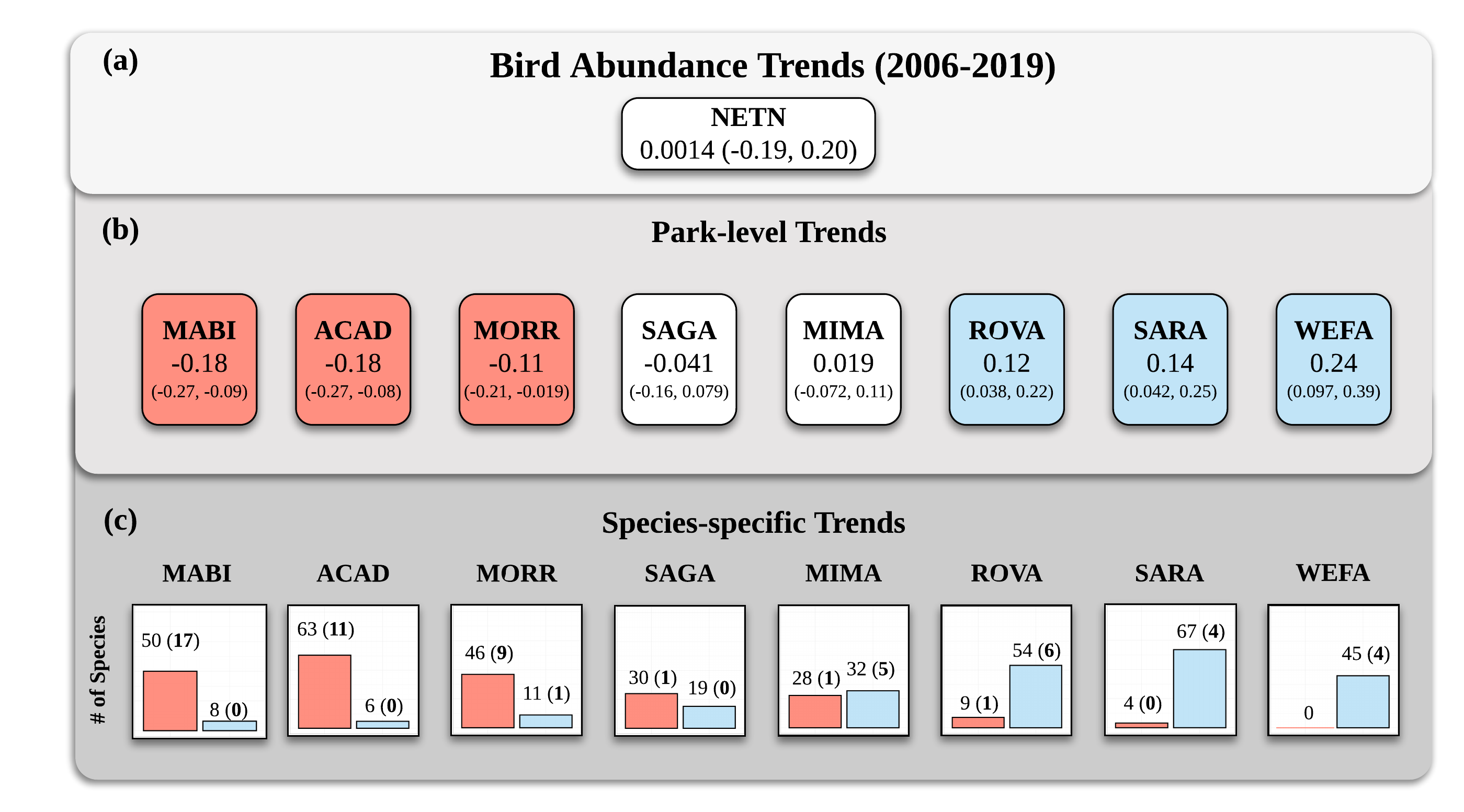}
    \caption{}
    \label{fig:trendDiagram}
\end{figure}

\begin{figure}
    \centering
    \includegraphics[width = 15cm]{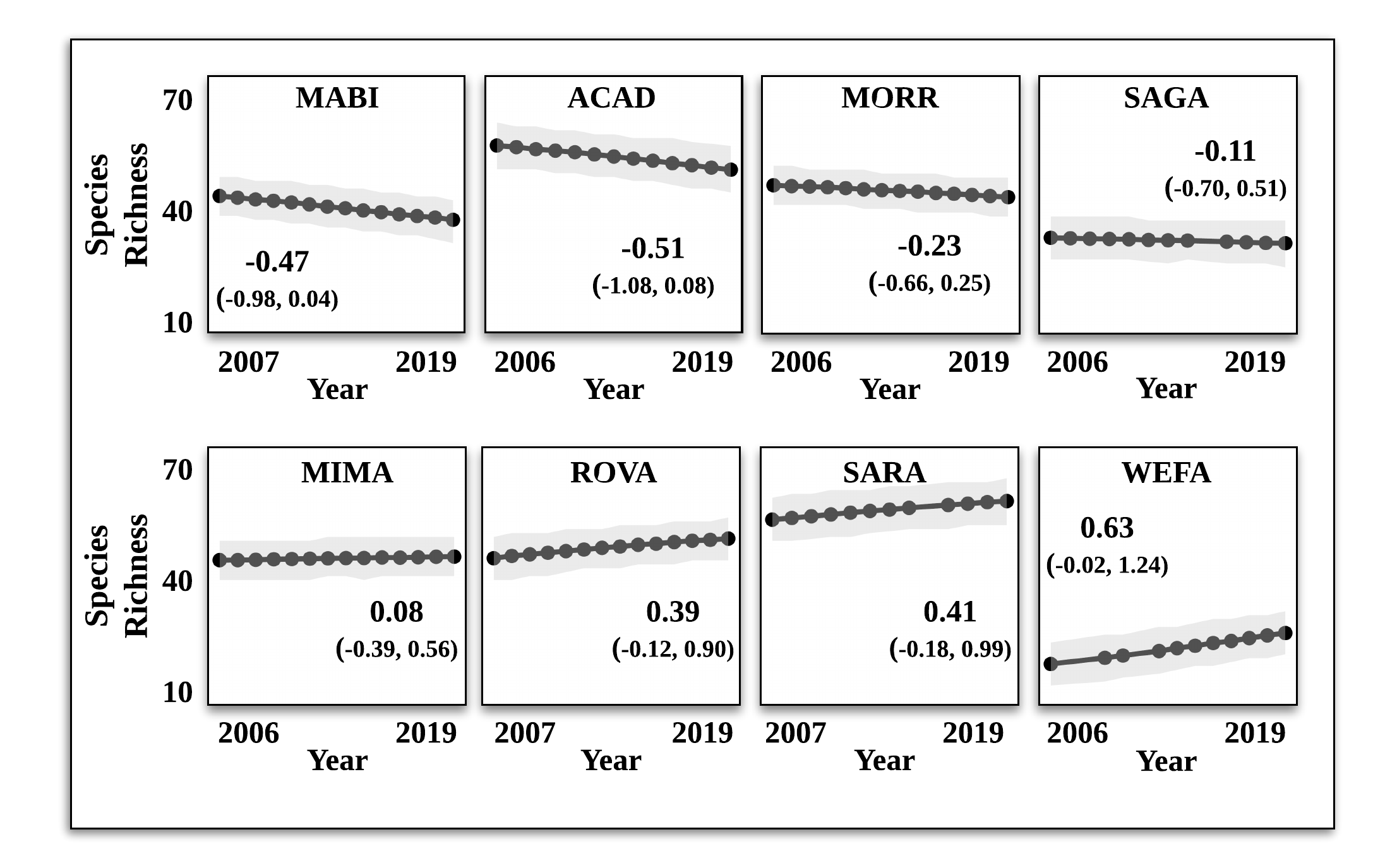}
    \caption{}
    \label{fig:richnessDiagram}
\end{figure}

There was wide variation in species abundance trends among the eight parks with three parks (MABI, ACAD, MORR) showing significant declines in overall abundance (across all species), two parks (SAGA, MIMA) showing no trend, and three parks (ROVA, SARA, WEFA) showing significant increases in abundance over the time frame of the study (Figure~\ref{fig:trendDiagram}b). Park-level species richness showed similar trends to that of abundance (Figure ~\ref{fig:richnessDiagram}). Interestingly, the wide variation across parks resulted in no temporal trend in total bird abundance across the entire network (Figure~\ref{fig:trendDiagram}a), revealing the benefit of the hierarchical modeling approach to evaluating park-specific dynamics that might be masked at the overall network level (i.e., Simpson's paradox). To assess if the variability across spatial units was a result of bird community composition, we computed the Bird Community Index (BCI; \citealt{OConnell2000}), a measure of bird communities that provides inference on the ecological integrity of a site (See Appendix S1 for details of BCI computation). We performed a correlation analysis between the estimated year trend and the mean estimated BCI, which revealed a strong negative correlation (median = -0.77).

Individual species abundance trends were highly variable across parks, but species within a given park tended to show similar trends (Figures~\ref{fig:trendDiagram}c, Appendix S1: Figure S1). Further, species in the same guild showed similar trends in abundance within a given park. However, the direction and magnitude of these trends varied widely across parks (Figure~\ref{fig:guildSummary}).

\subsection*{\textit{Forest structure effects}}

\begin{figure}
    \centering
    \includegraphics[width=15cm]{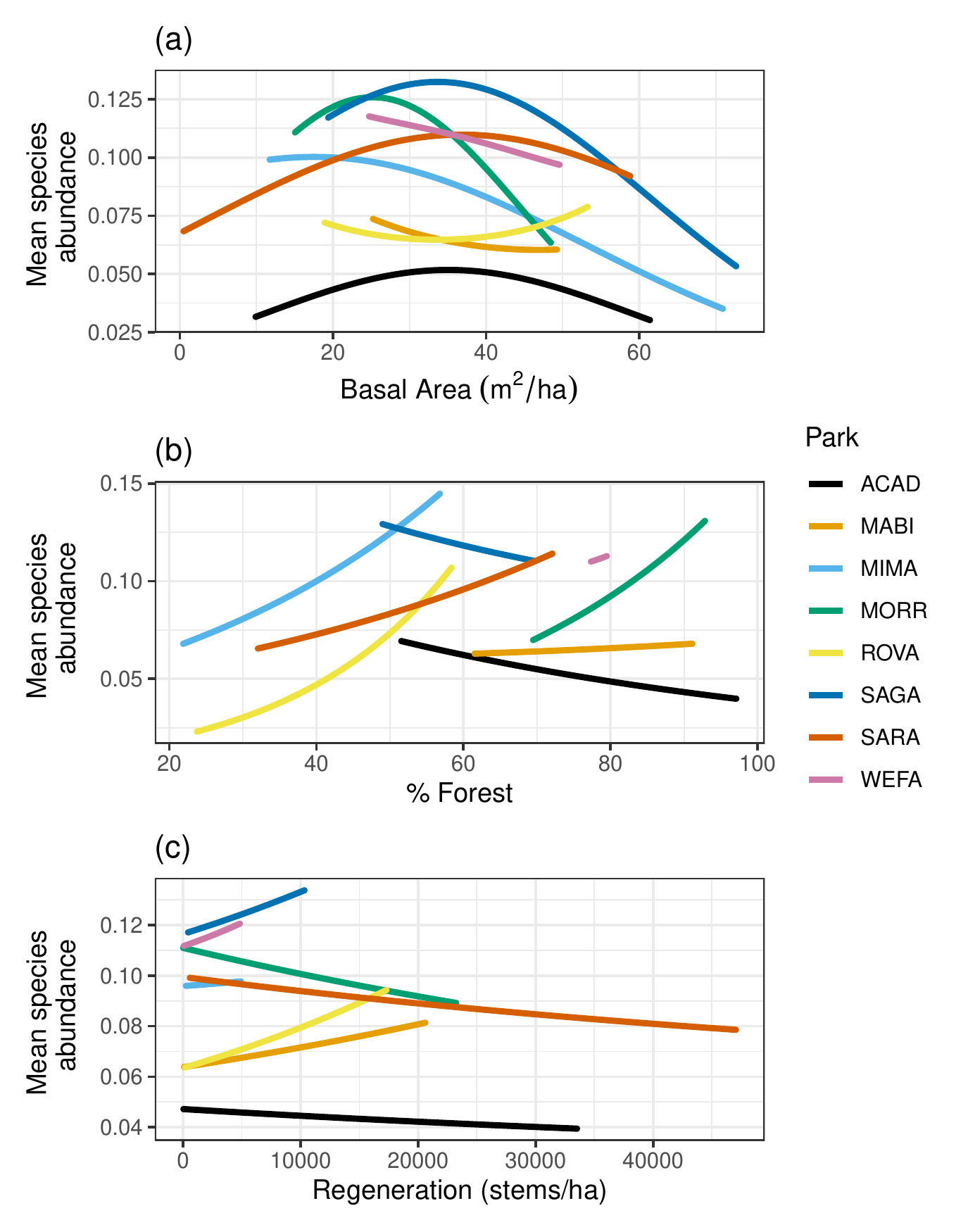}
    \caption{}
    \label{fig:covAbundanceEffects}
\end{figure}

At the network level, we found moderate support for a peak in bird abundance at intermediate values of basal area (85\% and 78\% probability of significant linear and quadratic effects, respectively), high amounts of local forest cover (92\% probability of a positive effect), and to a lesser extent high amounts of forest regeneration (70\% probability of a positive effect). Most parks followed the network level pattern, although there was variation among parks, especially in the effects of percent forest and forest regeneration (Figure~\ref{fig:covAbundanceEffects}). Percent forest has a more prominent effect than either basal area or forest regeneration, as evidenced by comparison of the magnitudes of the standardized regression coefficients. There was little variability in the effects of the covariates across different bird guilds (Appendix S1: Table S1).

\section*{Discussion}

With the growing threats of habitat loss, invasive species, and climate change, large-scale monitoring networks are becoming increasingly important for monitoring trends in species distributions and abundance across space and time. Such large-scale programs are often limited by resources at certain locations, indicating a need for sophisticated modeling approaches to estimate species trends across the entire region of interest. Our novel hierarchical model reveals large variations in bird abundance trends across eight protected forests but not across bird guilds within a park, suggesting that ecological processes, biological invasions, and management activities that affect local forest condition appear to have consistent effects on local forest bird communities. An understanding of how these variables influence diverse communities of bird species is crucial to informed and lasting management solutions for both forests and birds. 

\begin{figure}
    \centering
    \includegraphics[width = 15cm]{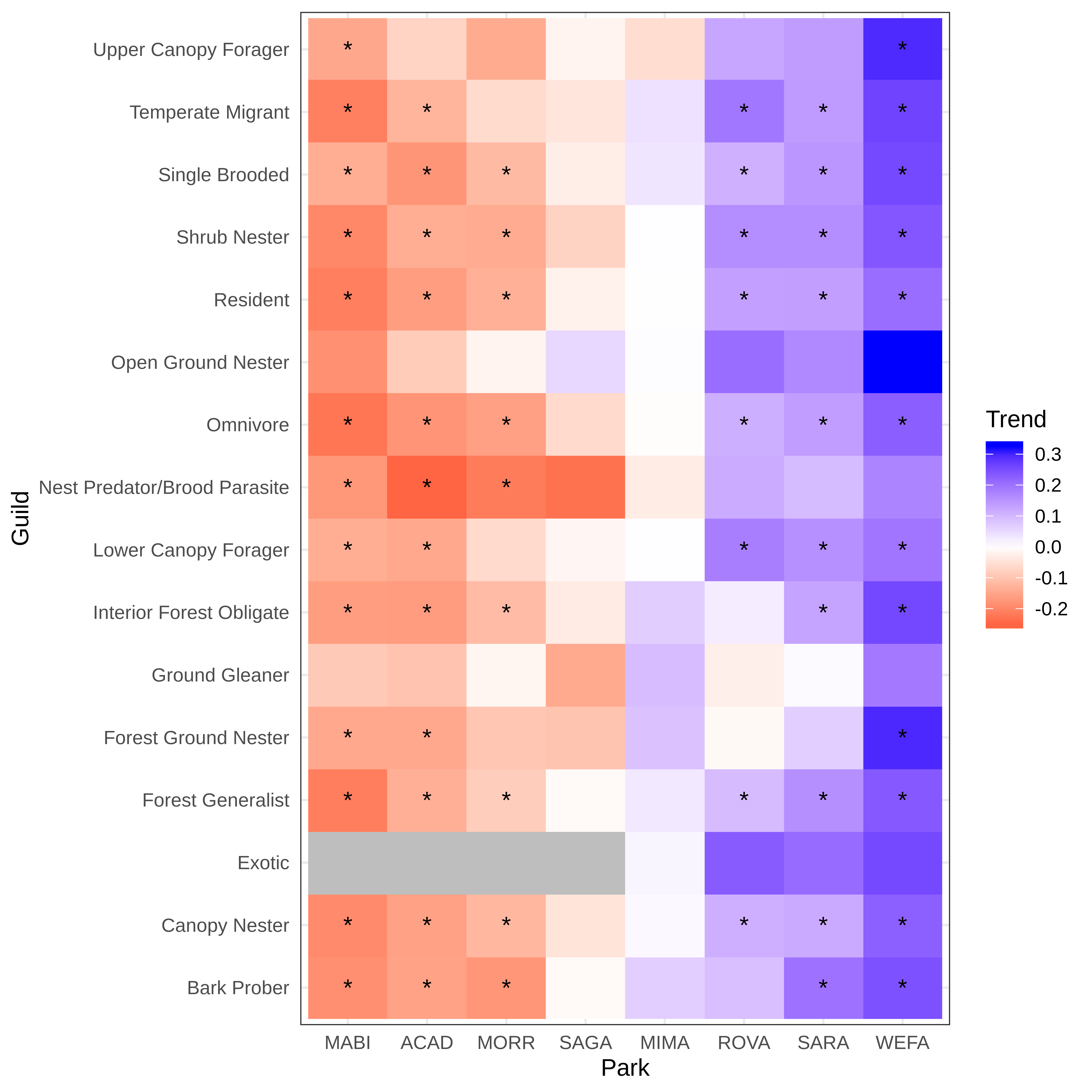}
    \caption{}
    \label{fig:guildSummary}
\end{figure}

Trends in bird abundance differed across space (i.e., parks) but not by species guilds (Figure~\ref{fig:guildSummary}), suggesting that local forest conditions might have broad and consistent effects on bird communities within parks. The consistent trends in species across guilds within a park are surprising given the variable responses bird guilds show towards insect disturbances \citep{Janousek2019}, silviculture \citep{Thiollay1997, Augenfeld2008}, and elevation \citep{Tenan2017}. However, life-history characteristics, including diet, foraging strategy, habitat preference, and nesting location, do not predict the effects of climate change on bird species distributions in northeastern North America \citep{Zuckerberg2009, Cohen2020}, suggesting responses of bird species to stressors such as climate change might be independent of life-history traits. The lack of variability we found in trends across guilds is partially attributable to species being assigned to multiple guilds. However, this cannot explain the variability in trends of guilds across the different parks (Figure~\ref{fig:guildSummary}). Thus, our results suggest that diverse communities of forest birds might show similar responses to variations in local forest condition and structure, as well as other spatially-dependent stressors, such as climate change \citep{Zuckerberg2009, Cohen2020}. 

We identified three parks (MABI, ACAD, MORR) with significant negative declines in bird abundance, two parks (MIMA, SAGA) with no trend, and three  parks (ROVA, SARA, WEFA) with significant, positive trends (Figure~\ref{fig:trendDiagram}). Understanding the causes of variability in bird abundance trends across parks and why species within parks behave fairly consistently is critical. Our findings correspond with previous studies showing variability in bird abundance and distribution trends across large spatial regions in the U.S. \citep{Rosenberg2019, Rushing2020}. The strong negative correlation between the BCI and the estimated year trend (median = -0.77) suggests bird communities reflective of higher ecological integrity are showing the fastest declines over time (Appendix S1: Figure S2). This result potentially suggests communities of birds within these parks might respond differently over time as a result of differences in local environmental stressors and interactions with other species \citep{hutchinson1957}, a phenomenon commonly addressed in species distribution models \citep{Pollock2014, Wilkinson2019, Lany2020}. 

While overall declines of bird abundance in MABI, ACAD, and MORR are concerning, it is important to emphasize that declines in each of these guilds should be viewed differently. Declines in specialist guilds (e.g., interior forest obligates, canopy nesters) are of highest concern, as these species are indicative of bird communities with high ecological integrity \citep{OConnell2000, Ladin2016}. Declines in generalist guilds (e.g., nest predators/brood parasites, forest generalists) are of lesser concern, and such declines can actually lead to an increase in the ecological integrity of the bird community as measured by the BCI. Thus, future analyses and management efforts should focus on declining specialist guilds in these three parks that require the interior and older forest habitat these parks are designed to protect. 

The amount of local forest cover has the largest effect on bird abundance across the parks (Table~\ref{tab:regionEffects}, Figure~\ref{fig:covAbundanceEffects}). Four parks (MORR, MIMA, ROVA, SARA) showed significant positive relationships between bird abundance and percent forest, which is consistent with our hypothesis and previous findings \citep{Willson1974, Ladin2016}. However, ACAD showed a significant negative relationship and SAGA showed a non-significant negative relationship, indicating higher abundance of birds at points surrounded by lower amounts of forest cover. The non-significant results at SAGA are likely a result of the low amount of variation in local forest cover (range of 49-70\%) near survey points. However, local forest cover at ACAD varied considerably among points (51\% to 97\%). The forests in ACAD and the surrounding forest matrix show distinct characteristics compared to the other seven parks in terms of their structural stage, density of large (>30 cm DBH) trees, and tree species diversity \citep{Miller2016, Miller2018}---underscoring that local forest cover alone does not account for all forest characteristics potentially important to bird breeding and foraging ecology. Further, it is important to note that while abundance in ACAD declines with forest cover, these sites were surrounded by relatively high forest cover, and even the points with the lowest forest cover are still surrounded by substantial forested habitat that could be enough habitat to attract and maintain many forest dwelling species \citep{Willson1974, Zuckerberg2010}. Overall, these results suggest management should focus on limiting forest fragmentation and maintaining or increasing the amount of forested cover in the surrounding landscape matrix, followed by maintenance of forest structure and diversity. However, the inconsistent effect of local forest cover across parks suggests that forest breeding birds are likely affected by interactions between local forest structure, surrounding land use, local community interactions, climate, and local stand dynamics (e.g., pest outbreaks, disturbances, succession). 

Our finding that overall abundance peaked at intermediate levels of live tree basal area in five park forests is consistent with previous research \citep{flaspohler2002, Rankin2015, VanderYacht2016}. This effect is likely related to larger variety in vertical forest structure at intermediate levels that provides suitable habitat to a wider variety of species \citep{macArthur1961, Crosby2020}. More broadly, the peaking of abundance at medium levels of live basal area could be reflective of multi-aged stands supporting a variety of species, individual tree size, and vertical layering that provides habitat to a wide range of birds.

Forest regeneration showed the weakest effect of the three covariates, with only a small positive relationship on overall bird abundance and significant at only one park (ROVA). Multiple parks had highly positive-skewed distributions of forest regeneration, which could indicate that there was not enough regeneration to attract birds that utilize such habitat. Given the legacy of deer overabundance, invasive species, and regional declines in regeneration that have shaped the understory in these parks \citep{Miller2019a}, more attention to examining the effects of regeneration and invasive plant abundance on forest birds seems warranted. Additionally, forest management aimed at increasing regeneration abundance and diversity within these forests should also be evaluated for its benefits to the overall bird community. 

While there was support for effects of tree basal area, forest regeneration, and local forest cover on bird abundance, these variables alone cannot explain the consistent trends of species within parks. ACAD, MABI, and MORR on average had the most forest within a 1km radius of each point. However, these forests vary considerably in structural stage \citep{Miller2016}, type, and diversity \citep{Miller2018}, suggesting that declines in bird abundance might be driven by factors other than forest structure. MABI is the only park in the network that is subject to logging, which has led to decreases in basal area, crown closure, and tree density and increases in regeneration over the study period that could contribute to the strong decreasing trends in bird abundance. MORR has experienced a decrease in crown closure as a result of intense storms and the invasive Emerald Ash Borer (\emph{Agrilus planipennis}), leading to increased gaps in the forests that could contribute to the decreasing trend in overall bird abundance. Reasons for the declines in ACAD are more ambiguous, but they could partially be attributed to declines in early successional habitat throughout the park. Overall, these variations in trends across spatial units are likely a composite result of numerous effects such as agricultural intensification and land use change outside of the parks \citep{Pino2000}, variations in forest age, structure, and size \citep{Miller2016}, differing levels of anthropogenic mortality, loss of sensitive areas or resources, migratory behavior, and climate change effects occurring during both the breeding and nonbreeding seasons \citep{Rushing2020}. 

Because the point counts did not contain specific information on the spatial location of each individual bird, we chose to describe population sizes in terms of point-level abundances rather than a measure of density, as the true effective sampling area of the point count is not well-defined due to movement and temporary immigration/emigration of birds \citep{kery2015applied}. This inability to accurately describe the sampling area is an inherent problem of any non-spatial method of counting birds (e.g., repeated counts, removal sampling) and is only remedied via approaches that incorporate spatial information of each individual, such as distance sampling \citep{buckland1993} and spatial capture-recapture \citep{efford2004, borchers2008spatially, royle2008hierarchical}. However, inference on point-level abundances, as used in the current study, provides information on population dynamics in time and space, which can yield management-relevant insights \citep{yamaura2011JAE, Chandler2013}. 

We developed a novel hierarchical model that enables inference on abundance for all 106 species observed across the eight parks and provides broad inference on how forest conditions affect bird species. We extended previous work on community abundance models \citep{yamaura2011JAE, yamaura2012BC, Chandler2013} to efficiently model communities of species across multiple geographically distinct regions. Our flexible modeling approach is widely applicable to monitoring programs where inference is desired on abundance of rare species and abundance of species at sparsely sampled locations. Further, our modeling approach can be easily extended to other sampling protocols commonly used to estimate abundance (e.g., distance sampling). Many of the 106 species in our data set are rare (i.e., lots of zeros in data), which leads to a trade-off between the number of species for which we can estimate trends and the number of covariates we can include in the analysis. This, in addition to the temporal misalignment of the forest covariates, limited our ability to use time-varying covariates to relate variability in forest structure over time to bird abundance trends. The addition of variables reflecting changes in species or vertical structure diversity (e.g., Shannon diversity, Gini diversity), as well as other variables related to climate change could likely provide additional insight on our results. Future work might also seek to combine additional data sources (e.g., Breeding Bird Survey, acoustic recordings, survival/productivity data) with our data using an integrated modeling approach (e.g., \citealt{Miller2019}) to help address this trade off, provide further inference on what is driving these trends in specialist guilds indicative of high integrity bird communities, and provide a more mechanistic approach at assessing population declines \citep{Zipkin2018, Saunders2019}.

\section*{Acknowledgements}

JWD, ASW, EFZ, and AOF conceived the modeling approach; ASW and KMM managed and organized the data collection; JWD analyzed the data and led writing of the manuscript. All authors  contributed critically to the drafts and gave final approval for publication. We declare that there are no conflicts of interest. We are grateful to Camilla Seirup and the field crews that contributed to the collection of NETN forest data in this study, to the volunteers, Steve Faccio, and Ed Sharron for their contributions to the NETN bird monitoring program, and to Adam Kozlowski and Brian Mitchell for their contributions to both programs. We thank John Marzluff, Marc K{\'e}ry, and an anonymous reviewer for insightful comments that greatly improved the manuscript. This work was supported by the National Park Service's Inventory and Monitoring Division and National Science Foundation grants DMS-1916395, EF-1253225, and DBI-1954406.

\section*{Open Research}

Data and code \citep{dataAvail} are available on Zenodo at: \url{https://doi.org/10.5281/zenodo.4701477}.

\bibliographystyle{apa}
\bibliography{references}

\newpage 

\section*{Tables}

\begin{table}[ht!] 
  \begin{center}
  \caption{General information on NETN parks included in the landbird monitoring program.}
  \label{tab:netnInfo}
  \begin{tabular}{c c c c c}
    \toprule
    Park Name & Code & Forested Area & Years Sampled & Points/Year \\
    & & (Ha) & (Year Established) & Mean (sd) \\
    \midrule
    Acadia & ACAD & 8178 & 13 (2007) & 35.8 (9.6)\\
    Saratoga & SARA & 687 & 12 (2007) & 13.3 (6.5) \\
    Morristown & MORR & 626 & 14 (2006) & 20.6 (6.4) \\
    Roosevelt-Vanderbilt & ROVA & 338 & 14 (2007) & 24.4 (6.1) \\
    Minute Man & MIMA & 234 & 14 (2006) & 21.5 (3.0) \\
    Marsh-Billings-Rockefeller & MABI & 196 & 14 (2006) & 23.6 (2.3)\\
    Saint-Gaudens & SAGA & 48 & 12 (2006) & 7.9 (0.3)\\
    Weir Farm & WEFA & 18 & 11 (2006) & 5 (0)\\
    \bottomrule
  \end{tabular}
  \end{center}
\end{table}

\newpage

\begin{table}[ht!] 
  \begin{center}
  \caption{Summary statistics of the observed bird community at each park in the NETN. Species abundance is the observed number of individuals of a given species at a given point.}
  \label{tab:summaryStats}
  \begin{tabular}{c c c c}
    \toprule
    Park & Observed Species & Average Species Abundance & Species Abundance Range\\
    & & (birds/point) & (birds/point) \\
    \midrule
    ACAD & 69 & 0.11 & [0.0021, 1.43]\\
    MABI & 58 & 0.20 & [0.0030, 2.95]\\
    MIMA & 60 & 0.22 & [0.0033, 1.46]\\
    MORR & 57 & 0.22 & [0.0035, 1.31]\\
    ROVA & 63 & 0.15 & [0.0029, 0.83]\\
    SAGA & 49 & 0.21 & [0.011, 1.79]\\
    SARA & 71 & 0.20 & [0.0063, 1.44]\\
    WEFA & 45 & 0.18 & [0.018, 1.16]\\
    \midrule
    NETN & 106 & 0.17 & [0.0021, 2.95]\\
    \bottomrule
  \end{tabular}
  \end{center}
\end{table}

\newpage 

\begin{table}[ht!] 
  \begin{center}
  \caption{Information on bird guilds used in hierarchical model. Adopted from \citet{OConnell2000}.}
  \label{tab:guildInfo}
  \begin{tabular}{c c c c}
    \toprule
    Biotic Integrity Element & Response Guild & Type & Number of Species \\
    \midrule
    Functional & Omnivore & Generalist & 34 \\
    Functional & Bark prober & Specialist & 10 \\
    Functional & Ground gleaner & Specialist & 7 \\
    Functional & Upper canopy forager & Specialist & 11 \\
    Functional & Lower canopy forager & Specialist & 20 \\
    Compositional & Exotic & Generalist & 4 \\
    Compositional & Resident & Generalist & 29 \\
    Compositional & Single-brooded & Specialist & 65 \\
    Compositional & Nest predator/brood parasite & Generalist & 7 \\
    Compositional & Temperate migrant & Generalist & 26 \\
    Structural & Canopy nester & Specialist &  31 \\
    Structural & Shrub nester & Generalist & 20 \\
    Structural & Forest-ground nester & Specialist & 14 \\
    Structural & Interior forest obligate & Specialist & 29 \\
    Structural & Forest generalist & Generalist & 25 \\
    Structural & Open-ground nester & Specialist & 9 \\
    \bottomrule
  \end{tabular}
  \end{center}
\end{table}

\newpage

\begin{table}[ht!] 
  \begin{center}
  \caption{Park-specific posterior medians of covariate effects on abundance and detection. Boldface indicates significance.}
  \label{tab:regionEffects}
  \begin{tabular}{c | c c c c c | c c c }
    \toprule
    Park & & & Abundance & & & & Detection & \\
    \midrule
     & BA & BA$^2$ & \%Forest & Regen & Year & Day & Day$^2$ & Time\\
    \midrule 
    MABI & \textbf{-0.16} & 0.075 & 0.050 & 0.093 & \textbf{-0.18} & 0.0013 & -0.087 & -0.16 \\
    ACAD & 0.042 & \textbf{-0.15} & \textbf{-0.25} & -0.046 & \textbf{-0.18} & 0.091 & -0.02 & -0.12 \\
    MORR & \textbf{-0.28} & \textbf{-0.24} & \textbf{0.54} & -0.077 & \textbf{-0.11} & \textbf{0.44} & -0.12 & -0.032\\
    SAGA & 0.009 & \textbf{-0.12} & -0.15 & 0.11 & -0.041 & 0.093 & -0.012 & 0.14\\
    MIMA & \textbf{-0.16} & \textbf{-0.069} & \textbf{0.45} & 0.025 & 0.019 & 0.11 & -0.041 & \textbf{-0.22} \\
    ROVA & -0.006 & 0.091 & \textbf{0.91} & \textbf{0.18} & \textbf{0.12} & 0.16 & -0.077 & \textbf{0.31} \\
    SARA & 0.037 & \textbf{-0.071} & \textbf{0.29} & -0.046 & \textbf{0.14} & -0.078 & -0.024 & -0.071 \\
    WEFA & -0.094 & -0.02 & 0.12 & 0.23 & \textbf{0.24} & -0.12 & -0.0045 & 0.035 \\
    \midrule
    NETN & -0.074 & -0.064 & 0.26 & 0.047 & 0.0014 & 0.088 & -0.047 & -0.012 \\
    \bottomrule
  \end{tabular}
  \end{center}
\end{table}

\newpage

\section*{Figure Legends}

\hspace{5mm} Figure~\ref{fig:studyArea}: Location of NETN parks participating in the breeding landbird monitoring program in the northeastern United States. See Table~\ref{tab:netnInfo} for description of park codes.

Figure~\ref{fig:modelDiagram}: Overview of hierarchical model components. Observations $\bm{y}_{r, i, j, t}$ are obtained for species $i$ in park $r$ at point $j$ in year $t$ using traditional point count surveys. Species level detection ($\bm{\alpha}_{r, i}$) and abundance ($\bm{\beta}_{r, i}$) parameters follow common park-level distributions with park level coefficients ($\bm{\mu}_{p, r}$, $\bm{\mu}_{\lambda, r}$), which subsequently follow common network level distributions with network level coefficients $\bar{\bm{\mu}}_{p}$ and $\bar{\bm{\mu}}_{\lambda}$, respectively.

Figure~\ref{fig:trendDiagram}: Trends in bird abundance from 2006-2019. Panel (a) shows the median linear trend of year across the entire network with the 95\% credible interval in parentheses. The trend estimates represent the average change in abundance on the log scale of a species per point every four years. Panel (b) shows the park level trends. Red highlight indicates a significant negative year trend, white highlight indicates no significant trend, and blue highlight indicates positive significant trend. Panel (c) shows the number of species with median linear trend of year estimates that are negative (red) and positive (blue) within each park. The number of species with significant trends is shown in boldface in parentheses.

Figure~\ref{fig:richnessDiagram}: Trends in estimated park-level species richness from 2006-2019 at each of the eight parks. Points are the mean species richness across all points sampled in each given park, gray regions denote the 95\% credible intervals. Inset text is the median (95\% credible interval) linear trend estimate for a post-hoc analysis between park-level species richness and year, representing the estimated change per year in species richness per park.

Figure~\ref{fig:guildSummary}: Mean linear year effect at each park averaged across all species defined within 16 bird guilds. * denotes significance. Gray color indicates no species in that guild were detected.

Figure~\ref{fig:covAbundanceEffects}: Relationship between mean expected number of birds of a single species per point and basal area (a), percent forest in a 1km radius (b), and forest regeneration (c). Abundance estimates in each graph are estimated at the median levels of all other covariates within a given park.

\newpage 

\section*{Figures} 

\newpage 

\newpage

\clearpage 

\newpage 
\clearpage

\newpage

\newpage

\end{document}